\documentclass[10pt,a4paper]{article}
\usepackage[margin=1.25in]{geometry}
\usepackage[utf8]{inputenc}
\usepackage{amsthm,amsmath}
\usepackage{amsfonts}
\usepackage{amssymb}
\usepackage{hyperref}
\usepackage[utf8]{inputenc} 
\usepackage{graphicx}
\usepackage{array}
\usepackage{authblk}

\newcolumntype{y}[1]{>{\centering\let\newline\\\arraybackslash\hspace{0pt}}m{#1}}

\newcommand{\chartWidthScale}{1.2} 

\title{Enhancement and Recognition of Reverberant and Noisy Speech by Extending Its Coherence}

\begin{document}

\author[1]{Scott Wisdom\thanks{swisdom@uw.edu}}
\author[1]{Thomas Powers}
\author[1]{Les Atlas}
\author[1,2]{James Pitton}
\affil[1]{Department of Electrical Engineering, University of Washington}
\affil[2]{Applied Physics Laboratory, University of Washington}

\maketitle

\begin{abstract} 
Most speech enhancement algorithms make use of the short-time Fourier transform (STFT), which is a simple and flexible time-frequency decomposition that estimates the short-time spectrum of a signal. However, the duration of short STFT frames are inherently limited by the nonstationarity of speech signals. The main contribution of this paper is a demonstration of speech enhancement and automatic speech recognition in the presence of reverberation and noise by extending the length of analysis windows. We accomplish this extension by performing enhancement in the short-time fan-chirp transform (STFChT) domain, an overcomplete time-frequency representation that is coherent with speech signals over longer analysis window durations than the STFT. This extended coherence is gained by using a linear model of fundamental frequency variation of voiced speech signals. Our approach centers around using a single-channel minimum mean-square error log-spectral amplitude (MMSE-LSA) estimator proposed by Habets, which scales coefficients in a time-frequency domain to suppress noise and reverberation. In the case of multiple microphones, we preprocess the data with either a minimum variance distortionless response (MVDR) beamformer, or a delay-and-sum beamformer (DSB). We evaluate our algorithm on both speech enhancement and recognition tasks for the REVERB challenge dataset. Compared to the same processing done in the STFT domain, our approach achieves significant improvement in terms of objective enhancement metrics (including PESQ---the ITU-T standard measurement for speech quality). In terms of automatic speech recognition (ASR) performance as measured by word error rate (WER), our experiments indicate that the STFT with a long window is more effective for ASR.
\end{abstract}

\section{Introduction}

Enhancement and recognition of speech signals in the presence of reverberation and noise remains a challenging problem in many applications. Many past methods are prone to generating artifacts in the enhanced speech, and must trade off noise reduction against speech distortion. Recent approaches have started to address this issue, demonstrating improvements in both objective speech quality and automatic speech recognition \cite{yoshioka_impact_2014,delcroix_linear_2014}.

In this paper, we propose using a new time-frequency domain that is more coherent with speech signals over an extended period of time, which allows longer analysis windows. In turn, longer analysis windows provide a more narrowband spectral representation, which concentrates signal energy into smaller numbers of FFT bins. Within these bins, the signal-to-noise ratio (SNR) is increased, which results in less oversuppression of speech.
We combine a statistically optimal single-channel enhancement algorithm that suppresses background noise and reverberation with an adaptive time-frequency transform domain that is coherent with speech signals over longer durations than the short-time Fourier transform (STFT). Thus, we are able to use longer analysis windows while still satisfying the assumptions of the optimal single-channel enhancement filter. Multichannel processing is made possible using a classic minimum variance distortionless response (MVDR) beamformer or, in the case of two-channel data, a delay-and-sum beamformer (DSB) preceding the single-channel enhancement.

First, we review the speech enhancement and dereverberation problem, as well as the enhancement algorithm we use proposed by Habets \cite{habets_speech_2010}, which suppresses both noise and late reverberation based on a statistical model of reverberation (originally proposed by Lebart et al. \cite{lebart_new_2001}). Then, we describe the fan-chirp transform, proposed by Weruaga and K\`{e}pesi \cite{kepesi_adaptive_2006, weruaga_fan-chirp_2007} and improved upon by Cancela et al. \cite{cancela_fan_2010}, which provides an enhancement domain, the short-time fan-chirp transform (STFChT), that better matches time-varying harmonic content of voiced speech.

We discuss why performing the enhancement in the STFChT domain gives superior results compared to the STFT domain. Further improvements over our original submission \cite{wisdom_enhancement_2014} to the REVERB challenge \cite{kinoshita_reverb_2013} are described, and we explore more optimal parameter settings. We present both speech enhancement and recognition results on the REVERB challenge dataset \cite{kinoshita_reverb_2013}, which shows that our new method achieves superior results versus conventional STFT-based processing in terms of objective speech enhancement measures. Through our automatic speech recognition (ASR) experiments, we discover that STFT-based processing with a longer window results in the lowest word error rates. Thus, our algorithm is an example of an operation that improves enhancement and objective quality metrics, but for reasons we hypothesize the operation does not improve ASR. However, our enhancement method may be able to provide complementary features to conventional STFT-based processing.

Our basic multichannel (given multiple microphones) architecture of single-channel enhancement preceded by beamforming is not unprecedented. Gannot and Cohen \cite{gannot_speech_2004} used a similar architecture for noise reduction that consists of a generalized sidelobe cancellation (GSC) beamformer followed by a single-channel post-filter. Maas et al. \cite{maas_application_2012} employed a similar single-channel enhancement algorithm for reverberation suppression and observed promising speech recognition performance in even highly reverberant environments.

There have been several dereverberation and enhancement approaches that estimate and leverage the time-varying fundamental frequency $f_0$ of speech. Nakatani et al. \cite{nakatani_harmonicity-based_2007} proposed a dereverberation method using inverse filtering that exploits the harmonicity of speech to build an adaptive comb filter. Kawahara et al. \cite{kawahara_restructuring_1999} used adaptive spectral analysis and estimates of $f_0$ to perform manipulation of speech characteristics.

Droppo and Acero \cite{droppo_fine_2007} observed how the fundamental frequency of speech can change within an analysis window, and proposed a new framework that could better predict the energy of voiced speech. Dunn and Quatieri \cite{dunn_sinewave_2007} used the fan-chirp transform for sinusoidal analysis and synthesis of speech, and Dunn et al. \cite{dunn_sinewave_2009} also examined the effect of various interpolation methods on reconstruction error. Pantazis et al. \cite{pantazis_adaptive_2011} proposed an analysis/synthesis domain that uses estimates of instantaneous frequency to decompose speech into quasi-harmonic AM-FM components. Degottex and Stylianou \cite{degottex_analysis_2013} proposed another analysis/synthesis scheme for speech using an adaptive harmonic model that they claim is more flexible than the fan-chirp, as it allows nonlinear frequency trajectories.

Wisdom et al. showed that the fan-chirp transform can be used to build optimal detectors for nonstationary harmonics \cite{wisdom_extending_2014-1} and harmonically-modulated stationary processes with time-varying modulation frequency \cite{wisdom_extending_2014}. A preliminary version of this algorithm appeared in our REVERB challenge workshop paper \cite{wisdom_enhancement_2014}. To our knowledge, these recent papers are the first to use the fan-chirp transform for statistical signal processing.

\section{Background}

This section gives necessary background on single-channel suppression of noise and late reverberation and on the window duration- and hence coherence-extending fan-chirp transform.

\subsection{Optimal single-channel suppression of noise and late reverberation}

In this section, we review the speech enhancement problem and a popular statistical speech enhancement algorithm, the minimum mean-square error log-spectral amplitude (MMSE-LSA) estimator, which was originally proposed by Ephraim and Malah \cite{ephraim_speech_1984,ephraim_speech_1985} and later improved by Cohen \cite{cohen_optimal_2002}. We review the application of MMSE-LSA to both noise reduction and joint dereverberation and noise reduction. Joint dereverberation and noise reduction was proposed by Habets \cite{habets_speech_2010}).

\subsubsection{Noise reduction using MMSE-LSA}

A classic speech enhancement algorithm is the minimum mean-square error (MMSE) short-time spectral amplitude estimator proposed by Ephraim and Malah \cite{ephraim_speech_1984}. They later refined the estimator to minimize the MSE of the log-spectra \cite{ephraim_speech_1985}. We will refer to this algorithm as LSA (log-spectral amplitude). Minimizing the MSE of the log-spectra was found to provide better enhanced output because log-spectra are more perceptually meaningful. Cohen \cite{cohen_optimal_2002} suggested improvements to Ephraim and Malah's algorithm, which he referred to as ``optimal modified log-spectral amplitude" (OM-LSA).

Given samples of a noisy speech signal
\begin{equation}
y[n]=s[n]+v[n],
\end{equation}
where $s[n]$ is the clean speech signal and $v[n]$ is additive noise, the goal of an enhancement algorithm is to estimate $s[n]$ from the noisy observations $y[n]$. Clean speech and noise are additive in the STFT domain:
\begin{equation}
Y(d,k)=S(d,k)+V(d,k).
\end{equation}
The LSA estimator yields an estimate $\hat{A}(d,k)$ of the clean STFT magnitudes $|S(d,k)|$ (where $S(d,k)$ are assumed to have a proper complex-valued Gaussian distribution) by applying a frequency-dependent gain $G_\mathrm{LSA}(d,k)$ to the noisy STFT magnitudes $|Y(d,k)|$:
\begin{equation}
\hat{A}(d,k) = G_\mathrm{LSA}(d,k) |Y(d,k)|.
\end{equation}
Given these estimated magnitudes, the enhanced speech is reconstructed from STFT coefficients combining $\hat{A}(d,k)$ with noisy phase: 
\begin{equation}
\hat{S}(d,k)=\hat{A}(d,k)e^{j\angle Y(d,k)}.
\end{equation}
The LSA gains are computed as \cite[equation (20)]{ephraim_speech_1985}:
\begin{equation} \label{eq:lsa_em}
G_\mathrm{LSA}(d,k) = \frac{\xi(d,k)}{1+\xi(d,k)} \exp\left\{ \frac{1}{2} \int_{v(d,k)}^\infty \frac{e^{-t}}{t}dt \right\}.
\end{equation}
The lower integral bound in (\ref{eq:lsa_em}) is 
\begin{equation} \label{eq:lsa_nu}
v(d,k)=\frac{\xi(d,k)}{1+\xi(d,k)}\gamma(d,k),
\end{equation}
where $\xi(d,k)$ and $\gamma(d,k)$ are the {\it a priori} and {\it a posteriori} signal-to-noise ratios (SNRs), respectively, for the $k$th frequency bin of the $d$th frame. These SNRs are defined to be
\begin{equation}
\xi(d,k) \overset{\Delta}{=} \frac{\lambda_s(d,k)}{\lambda_v(d,k)}
\textrm{\;\; and \;\;} 
\gamma(d,k) \overset{\Delta}{=} \frac{|Y(d,k)|^2}{\lambda_v(d,k)}
\end{equation}
where 
\begin{equation}
\lambda_s(d,k)=E\left\{ |S(d,k)|^2 \right\}
\end{equation}
end
\begin{equation}
\lambda_v(d,k)=E\left\{ |V(d,k)|^2 \right\}
\end{equation}
are the variances of $S(d,k)$ and $V(d,k)$, respectively.

Cohen \cite{cohen_optimal_2002} refined Ephraim and Malah's approach to include a lower bound $G_\mathrm{min}$ for the gains as well as an {\it a priori} speech presence probability (SPP) estimator $p(d,k)$. Cohen's estimator is as follows \cite[equation (8)]{cohen_optimal_2002}:
\begin{equation} \label{eq:om_lsa}
G_\mathrm{OM-LSA}=\left\{ G_\mathrm{LSA}(d,k) \right\}^{p(d,k)}\cdot G_\mathrm{min}^{1-p(d,k)}.
\end{equation}
Cohen also derived an efficient estimator for the SPP $p(d,k)$ \cite{cohen_optimal_2002} that exploits the strong interframe and interfrequency correlation of speech in the STFT domain.

\subsubsection{Joint dereverberation and noise reduction} \label{ssec:mmse_lsa}

This subsection reviews a MMSE-LSA enhancement algorithm proposed by Habets \cite{habets_speech_2010} that uses a statistical model of reverberation to suppress both noise and late reverberation. Such a statistical model-based approach to dereverberation was originally proposed by Lebart et al. \cite{lebart_new_2001}. We will refer to this type of MMSE-LSA as HMMSE-LSA (for Habets MMSE-LSA). The signal model Habets uses is
\begin{equation}
y[n] = s[n]*h[n] + v[n] = x_e[n]+x_\ell[n]+v[n],
\end{equation}
where $s[n]$ is the clean speech signal, $h[n]$ is the room impulse response (RIR), and $v[n]$ is additive noise. The terms $x_e[n]$ and $x_\ell[n]$ correspond to the early and late reverberated speech signals, respectively. The partition between early and late reverberations is determined by a parameter $n_e$, which is a discrete sample index. All samples in the RIR before $n_e$ are taken to cause early reflections, and all samples after $n_e$ are taken to cause late reflections \cite{habets_speech_2010}. Thus,
\begin{equation} \label{eq:h}
h[n]= 
\begin{cases}
    0,				& \text{if } n<0\\
    h_e[n],    & \text{if } 0 \leq n < n_e\\
    h_\ell[n]	& \text{if } n_e \leq n.
\end{cases}
\end{equation}
Using these definitions, $x_e[n]=s[n]*h_e[n]$ and $x_\ell[n]=s[n]*h_\ell[n]$.

Habets proposed a generalized statistical model of reverberation that is valid both when the source-microphone distance is less than or greater than the critical distance \cite{habets_speech_2010}. This model divides the RIR $h[n]$ into a direct-path component $h_d[n]$ and reverberant component $h_r[n]$. Both direct-path and reverberant components are taken to be white, zero-mean, stationary Gaussian noise sequences $b_d[n]$ and $b_r[n]$ with variances $\sigma_d^2$ and $\sigma_r^2$ scaled by an exponential decay,
\begin{align} \label{eq:hd}
h_d[n] = b_d[n]e^{-\bar{\zeta}n} \;\;\mathrm{and}\;\; h_r[n] = b_r[n]e^{-\bar{\zeta}n},
\end{align}
where $\bar{\zeta}$ is related to the reverberation time $T_{60}$ by \cite{habets_speech_2010}:
\begin{equation} \label{eq:zeta}
\bar{\zeta} = \frac{3 \ln(10)}{T_{60}f_s}.
\end{equation}

Using this model, the expected value of the energy envelope of $h[n]$ is
\begin{equation} \label{eq:Eh2}
E\left[ h^2[n] \right] = 
\begin{cases}
    \sigma_d^2e^{-2\bar{\zeta}n},				& \text{for } 0\leq n<n_d\\
    \sigma_r^2e^{-2\bar{\zeta}n},    & \text{for } n \geq n_d \\
    0	& \text{otherwise, }
\end{cases}
\end{equation}
where $n_d$ is a parameter chosen to be the number of samples that correspond to the direct part of a reverberant signal.

\begin{figure}[h!]
 \centering
 \includegraphics[width=0.5\linewidth]{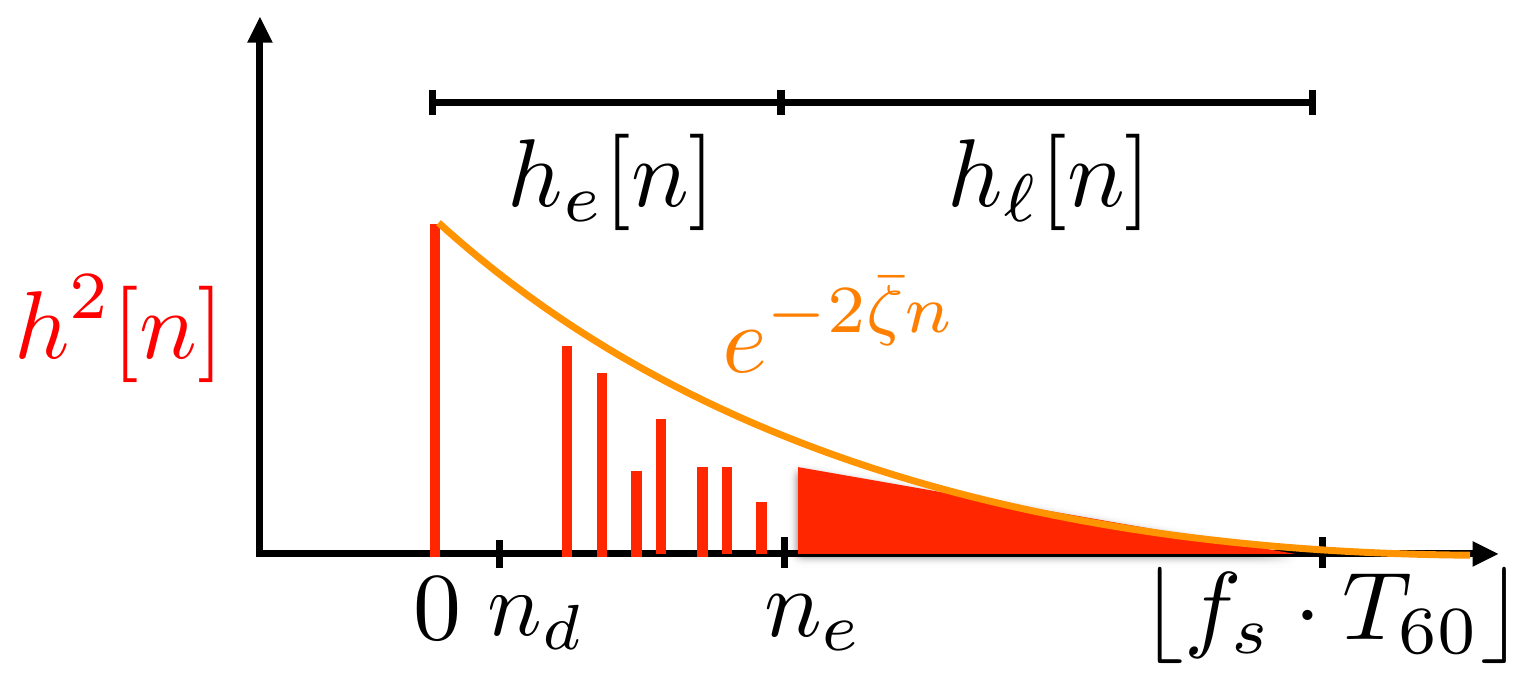}
  \caption{{\bf Reverberation model.}
      A schematic illustration of the statistical reverberation model given by equations (\ref{eq:h})-(\ref{eq:Eh2}).
      } \label{fig:reverbSch}
      \end{figure}

Figure \ref{fig:reverbSch} shows a schematic illustration of this statistical model of reverberation. Under the assumption that the speech signal is stationary over short analysis windows (i.e., duration much less than $T_{60}$), Habets proposed \cite[equation (3.87)]{habets_speech_2010} the following model of the spectral variance of the reverberant component $x_r[n]$, which is denoted by $\lambda_{x_r}(d,k)$:
\begin{align} \label{eq:reverb_var}
\lambda_{x_r}(d,k) =& e^{-2\bar{\zeta}(k)R}\lambda_{x_r}(d-1,k) ...\nonumber \\
&+ \frac{E_r}{E_d}\left( 1-e^{-2\bar{\zeta}(k)R} \right) \lambda_{x_d}(d-1,k),
\end{align}
where $R$ is the number of samples separating two adjacent analysis frames and $E_r/E_d$ is the inverse of the direct-to-reverberant ratio (DRR). The quantities $E_r$ and $E_d$ are the energies of the reverberant and direct components of the signal, respectively. The DRR expresses the energy level of the direct signal referenced to the energy level of the reverberant part. Thus, the spectral variance of the reverberant component in the current frame $d$ is composed of scaled copies of the spectral variance of the reverberation and the spectral variance of the direct-path signal from the previous frame $d-1$.

Using this model, the variance of the late reverberant component can be expressed as \cite[equation (3.85)]{habets_speech_2010}:
\begin{equation}
\lambda_{x_\ell}(d,k) = e^{-2\bar{\zeta}(k)(n_e-R)}\lambda_{x_r}\left( d-\frac{n_e}{R}+1,k \right),
\end{equation}
which is quite useful in practice, because the variance of the late-reverberant component can be computed from the variance of the total reverberant component.

To suppress both noise and late reverberation, the {\it a priori} and {\it a posteriori} SNRs $\xi(d,k)$ and $\gamma(d,k)$ from the previous section become {\it a priori} and {\it a posteriori} signal-to-interference ratios (SIRs), given by \cite[equations (3.25-26)]{habets_speech_2010}:
\begin{equation} \label{eq:sir_xi}
\xi(d,k) = \frac{\lambda_{x_e}(d,k)}{\lambda_{x_\ell}(d,k)+\lambda_v(d,k)}
\end{equation}
and
\begin{equation} \label{eq:sir_gam}
\gamma(d,k) = \frac{|Y(d,k)|^2}{\lambda_{x_\ell}(d,k)+\lambda_v(d,k)}.
\end{equation}
The gains are computed by plugging the SIRs in (\ref{eq:sir_xi}) and (\ref{eq:sir_gam}) into (\ref{eq:lsa_em}) and (\ref{eq:lsa_nu}). Habets suggested an additional change to (\ref{eq:om_lsa}), which makes $G_\mathrm{min}$ time- and frequency-dependent. This is done because the interference of both noise and late reverberation is time-varying. The modification is \cite[equation (3.29)]{habets_speech_2010}
\begin{equation}
G_\mathrm{min}(d,k)=\frac{ G_{\mathrm{min},x_\ell}\hat{\lambda}_{x_\ell}(d,k) + G_{\mathrm{min},v}\hat{\lambda}_{v}(d,k) }{\hat{\lambda}_{x_\ell}(d,k) + \hat{\lambda}_{v}(d,k)}.
\end{equation}

Notice that two parameters in (\ref{eq:zeta}) and (\ref{eq:reverb_var}) are not known {\it a priori}; namely, $T_{60}$ and the DRR. These parameters must be blindly estimated from the data. For $T_{60}$ estimation, L\"{o}llmann et al. \cite{lollmann_improved_2010} propose a maximum-likelihood algorithm, which we found to be effective. As for the DRR, Habets suggests an online adaptive procedure \cite[\S 3.7.2]{habets_speech_2010}. This adaptive procedure constrains the DRR between $0$ and $1$ and assumes that the source is within the critical distance (i.e., the distance at which direct and reverberant energy are equal). This assumption prevents overestimation of the reverberant variance when the direct signal is active.

%

\subsection{Analysis using the forward fan-chirp transform}

In this section, we review the forward short-time fan-chirp transform (STFChT), which is used as the time-frequency analysis-synthesis domain for our enhancement algorithm. In section \ref{ssec:istfcht}, we describe our novel method of inverting the STFChT. 

We adopt the fan-chirp transform formulation used by Cancela et al. \cite{cancela_fan_2010}. The forward fan-chirp transform is defined as
\begin{equation} \label{eq:fc1}
X(f,\alpha) = \int x(t) \phi'_\alpha(t) e^{-j2\pi f \phi_\alpha(t)} dt
\end{equation}
where $\phi_\alpha(t)=\left(t+\frac{1}{2}\alpha t^2\right)$ and $\phi'_\alpha(t) = 1+\alpha t$. The variable $\alpha$ is an analysis chirp rate. The chirp rate $\alpha$ is a normalized chirp rate; that is, if the total bandwidth swept is $B$ Hertz over a time duration $T$ seconds, then $\alpha=\frac{B}{Tf}$. Using a change of variable $\tau = \phi_\alpha(t)$, (\ref{eq:fc1}) can be written as the Fourier transform of a time-warped signal:
\begin{equation} \label{eq:fc2}
X(f,\alpha) = \int_{-\infty}^\infty x(\phi_\alpha^{-1}(\tau)) e^{-j2\pi f \tau} d\tau.
\end{equation}

The short-time fan-chirp transform (STFChT) 
of $x(t)$ is defined as the fan-chirp transform of the $d$th short frame of $x(t)$:
\begin{equation} \label{eq:stfcht}
X_d(f,\hat{\alpha}_d) = \int_{-T_w/2}^{T_w/2} w(\tau) x_d(\phi_{\hat{\alpha}_d}^{-1}(\tau)) e^{-j2\pi f \tau}d\tau
\end{equation}
where $w(t)$ is an analysis window, $\hat{\alpha}_d$ (given by (\ref{eq:glogs})) is the analysis chirp rate for the $d$th frame, and $x_d(t)$ is the $d$th short frame of the input signal of duration $T$:
\begin{equation}
x_d(t)=\left\{
\begin{array}{c l}      
    x(t-dT_{hop}), & -T/2\leq t \leq T/2\\
    0, & \mathrm{otherwise}.
\end{array}\right.
\end{equation}

$T$ is the duration of the pre-warped short-time duration, $T_{hop}$ is the frame hop, $T_w$ is the post-warped short-time duration, and $w(t)$ is a $T_w$-long analysis window. The analysis window is applied after time-warping so as to avoid warping of the window, which can cause unpredictable smearing of the Fourier transform.

Implementing the fan-chirp transform as a time-warping followed by a Fourier transform allows efficient implementation, consisting simply as an interpolation of the signal followed by an FFT. In the implementation provided by Cancela et al. \cite{cancela_fan_2010}, the interpolation used in the forward fan-chirp transform is linear.

K\`{e}pesi and Weruaga \cite{kepesi_adaptive_2006} provide a method for determination of the analysis chirp rate $\alpha$ using the gathered log spectrum (GLogS). The GLogS is defined as the harmonically-gathered log-magnitude spectrum:
\begin{equation} \label{eq:rawglogs}
\rho(f_0,\alpha) = \frac{1}{N_h}\sum_{k=1}^{N_h} \ln \left| X(kf_0,\alpha) \right|
\end{equation}
where $N_h$ is the maximum number of harmonics that fit within the analysis bandwidth. That is, 
\begin{equation}
N_h = \left\lfloor \frac{f_s}{2f_0 \left(1 + \frac{1}{2} |\alpha| T_w \right)} \right\rfloor.
\end{equation}

Cancela et al. \cite{cancela_fan_2010} proposed several enhancements to the GLogS. First, they observed improved results by replacing $\ln|\cdot|$ with $\ln \left(1+ \gamma \left| \cdot  \right| \right)$. Cancela et al. note that this expression approximates a $p$-norm, with $0<p<1$, where lower values of $\gamma$ with $\gamma \geq 1$ approach the 1-norm, while higher values approaches the 0-norm. Cancela et al. note that $\gamma=10$ gave good results for their application.

%

Additionally, Cancela et al. propose modifications that suppress multiples and submultiples of the current $f_0$. Also, they propose normalizing the GLogS such that it has zero mean and unit variance. This is necessary because the variance of the GLogS increases with increasing fundamental frequency. For means and variances measured over all frames in a database, a polynomial fit is determined and the GLogS are compensated using these polynomial fits.

Let $\bar{\rho}_d(f_0,\alpha)$ be the GLogS of the $d$th frame with these enhancements applied. For practical implementation, finite sets $\mathcal{A}$ of candidate chirp rates and $\mathcal{F}_0$ of candidate fundamental frequencies are used, and the GLogS is exhaustively computed for every chirp rate in $\mathcal{A}$ and fundamental frequency in $\mathcal{F}_0$. The analysis chirp rate $\hat{\alpha}_d$ for the $d$th frame is thus found by
\begin{equation} \label{eq:glogs}
\hat{\alpha}_d = \underset{\alpha \in \mathcal{A}}{\mathrm{argmax}}\; \underset{f_0 \in \mathcal{F}_0}{\mathrm{max}} \; \bar{\rho}_d(f_0,\alpha).
\end{equation}

\subsection{Synthesis using the inverse fan-chirp transform} \label{ssec:istfcht}

Inverting the fan-chirp transform is a matter of reversing the steps used in the forward transform. Thus, the inverse fan-chirp transform for a short-time frame consists of an inverse Fourier transform, removal of the analysis window, and an inverse time-warping. The removal of the analysis window $w(t)$ from the $T_w$-long warped signal limits the choice of analysis windows to non-zero functions only, such as a Hamming window, so the window can be divided out. Also, since the warping is nonuniform, it is possible that the sampling interval between points may exceed the Nyquist sampling interval. To combat the potential for aliasing, the data should be oversampled before time-warping, which requires downsampling after undoing the time-warping.

The choice of post-warped duration $T_w$ and the method of interpolation used in the inverse time-warping affect the reconstruction error of the inverse fan-chirp transform. There is a trade-off between reconstruction performance and computational complexity, because interpolation error decreases as interpolation order increases. K\`{e}pesi and Weruaga \cite{weruaga_speech_2004} analyzed fan-chirp reconstruction error with respect to order of the time-warping interpolation and oversampling factor, and found that for cubic splines and an oversampling factor of 2, a signal-to-error ratio of over 30dB can be achieved. For our application, we choose an oversampling factor of 8 and cubic-spline interpolation.



\section{Proposed Method}

As discussed in the introduction, our main contribution is that we use the short-time fan-chirp transform as the analysis-synthesis domain for the HMMSE-LSA algorithm. In this section, we describe two aspects of our proposed method. First, we discuss the benefits of performing enhancement in the short-time fan-chirp domain. Next, we describe our method of iterative enhancement, which provides additional improvement to the processing. We go on to show how the parameters of iterative enhancement and analysis window duration affect our processing.

\subsection{Advantage of HMMSE-LSA in the Fan-Chirp Domain} \label{ssec:advantage}
Unlike a conventional Fourier transform, the fan-chirp transform captures intra-window frequency variation. As a result, the fan-chirp transform better matches the frequency content of a harmonic signal and concentrates the signal's energy into fewer bins. To illustrate this property, we perform a comparison of the local time-frequency SNRs of the STFChT and the STFT. Both transforms are applied to a simulated signal of two linear harmonic chirps in a simulated noisy and reverberant environment. The first chirp has a fundamental frequency varying from 200 Hz to 233 Hz, and the second chirp decreases from 250 Hz to 200 Hz. Both chirps last for 200 ms and have 20 harmonics. To simulate reverberation, we convolve the signal with a measured room impulse response (RIR) corresponding to the medium size room 2 far condition from the WSJCAM0 speech corpus from the 2014 REVERB challenge dataset \cite{kinoshita_reverb_2013}. Recorded air conditioning noise from the same room is added at 20 dB SNR.

Since we know the analytical form of the test signal, we know precisely which time-frequency bins contain direct signal. Convolving this known test signal with a measured RIR and adding actual recorded noise allows us to view the true local SNR in each time-frequency bin of the two transforms for realistic reverberation and additive noise. Given a time-frequency transform $S(d,k)$ of the direct signal, the time-frequency transform $X_r(d,k)$ of the reverberant signal, and the time-frequency transform $V(d,k)$ of the noise, we compute local SNR in a time-frequency bin $(d,k)$ as 
\begin{equation}
SNR_{local}(d,k)=\frac{|S(d,k)|^2}{|X_r(d,k)|^2+|V(d,k)|^2}.
\end{equation}
Thus, we can observe this oracle local SNR for bins containing direct signal, noise, and reverberation, and for bins containing only noise and reverberation.

\begin{figure}
\centering
\includegraphics[width=0.65\linewidth]{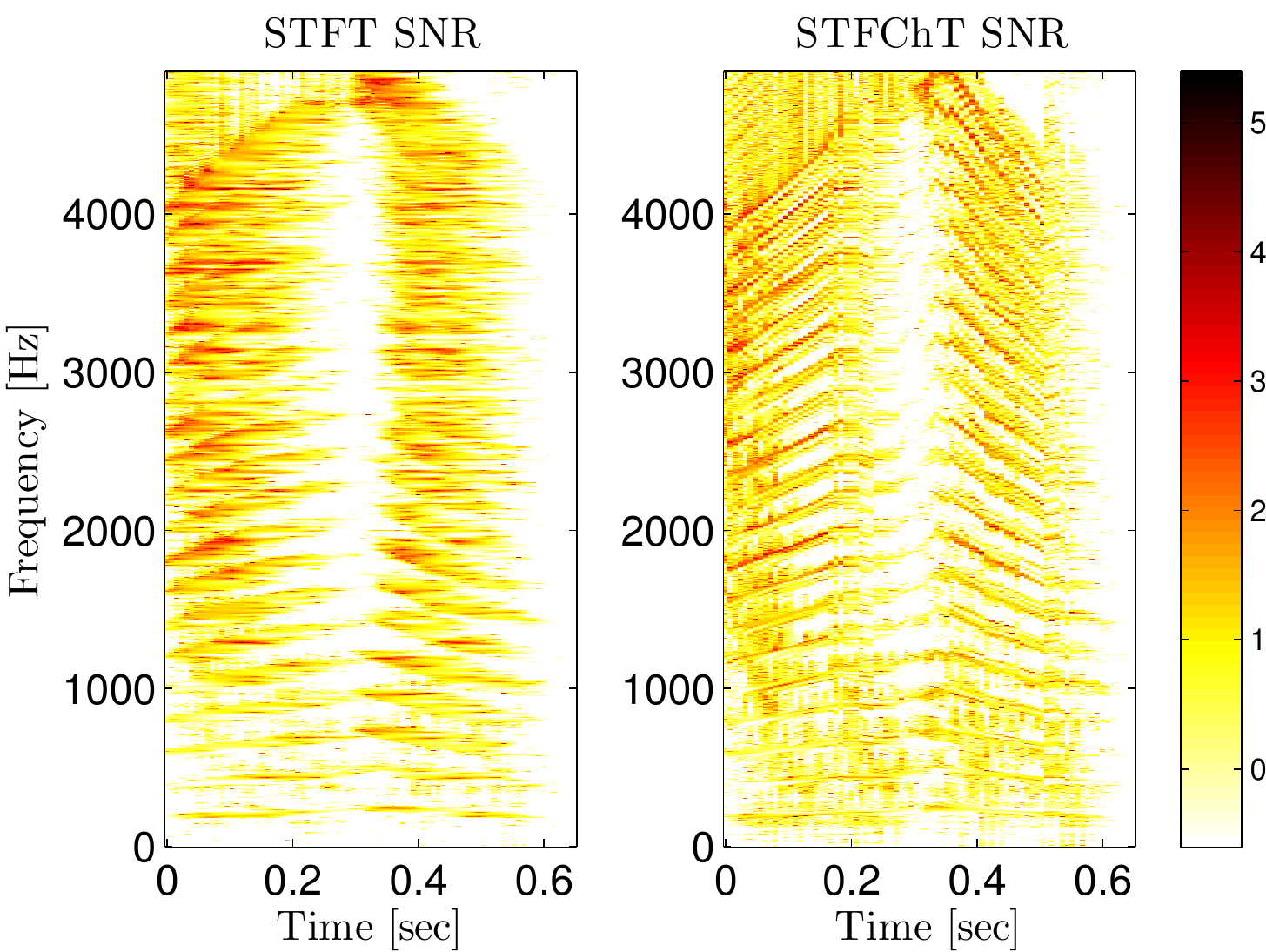}
\caption{{\bf Oracle local SNR values for a sequence of synthetic chirp signals}. These oracle local SNRs illustrate the less smeared concentration of SNR within individual time-frequency bins for direct path signal in the STFChT (right) as compared to the STFT (left).} \label{fig:snrmap}
\end{figure}

\begin{figure}[h]
\centering
\begin{minipage}[b]{0.45\linewidth}
	\includegraphics[width=\linewidth]{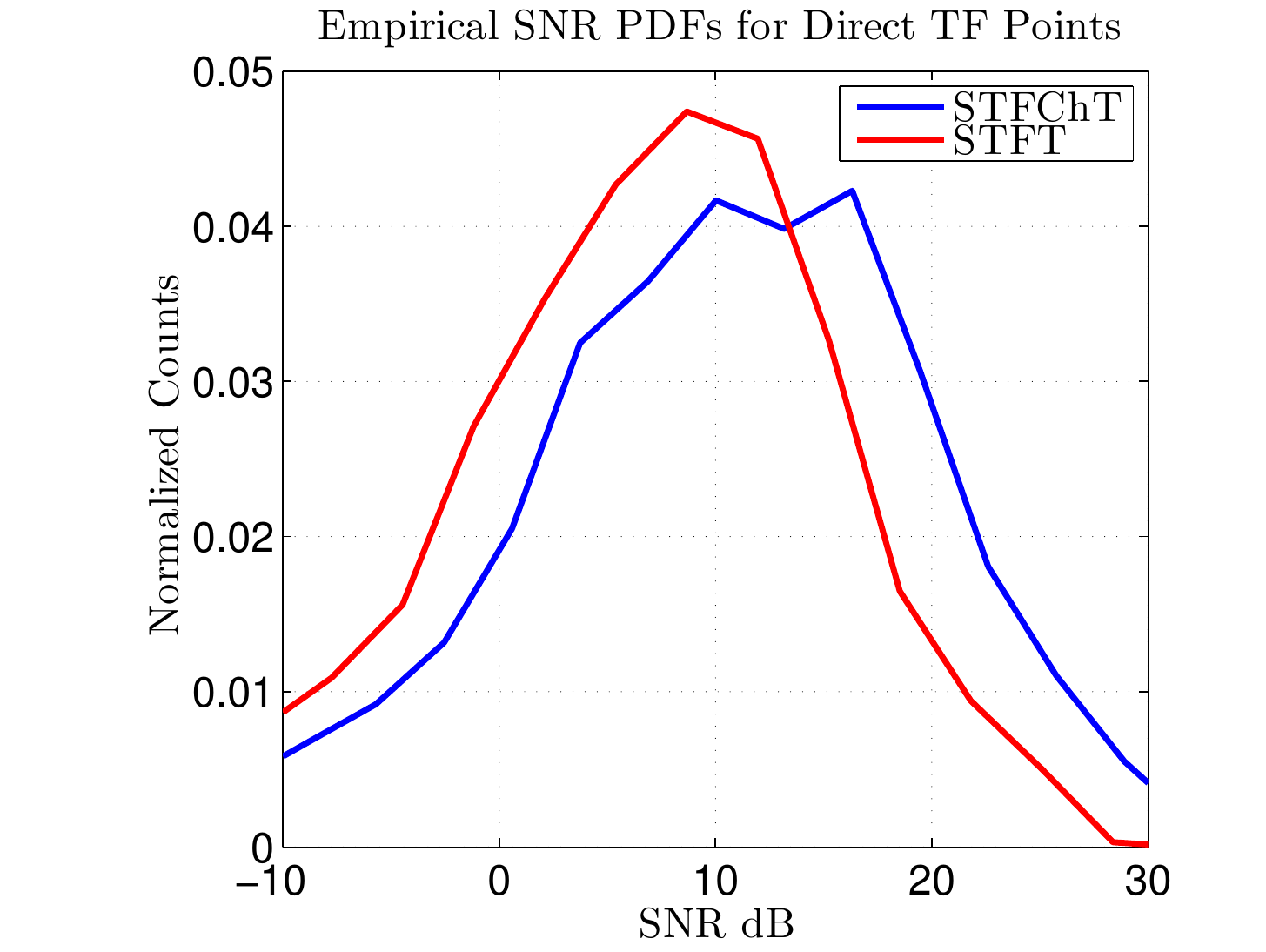}
	\caption{{\bf Empirical distribution of local SNR values in time-frequency bins containing direct signal.} Given for the synthetic chirp signals in figure \ref{fig:snrmap}. Notice that the STFChT provides a higher mean local SNR within time-frequency bins containing direct signal} \label{fig:histdir}
\end{minipage}
\hfill
\begin{minipage}[b]{0.45\linewidth}
	\includegraphics[width=\linewidth]{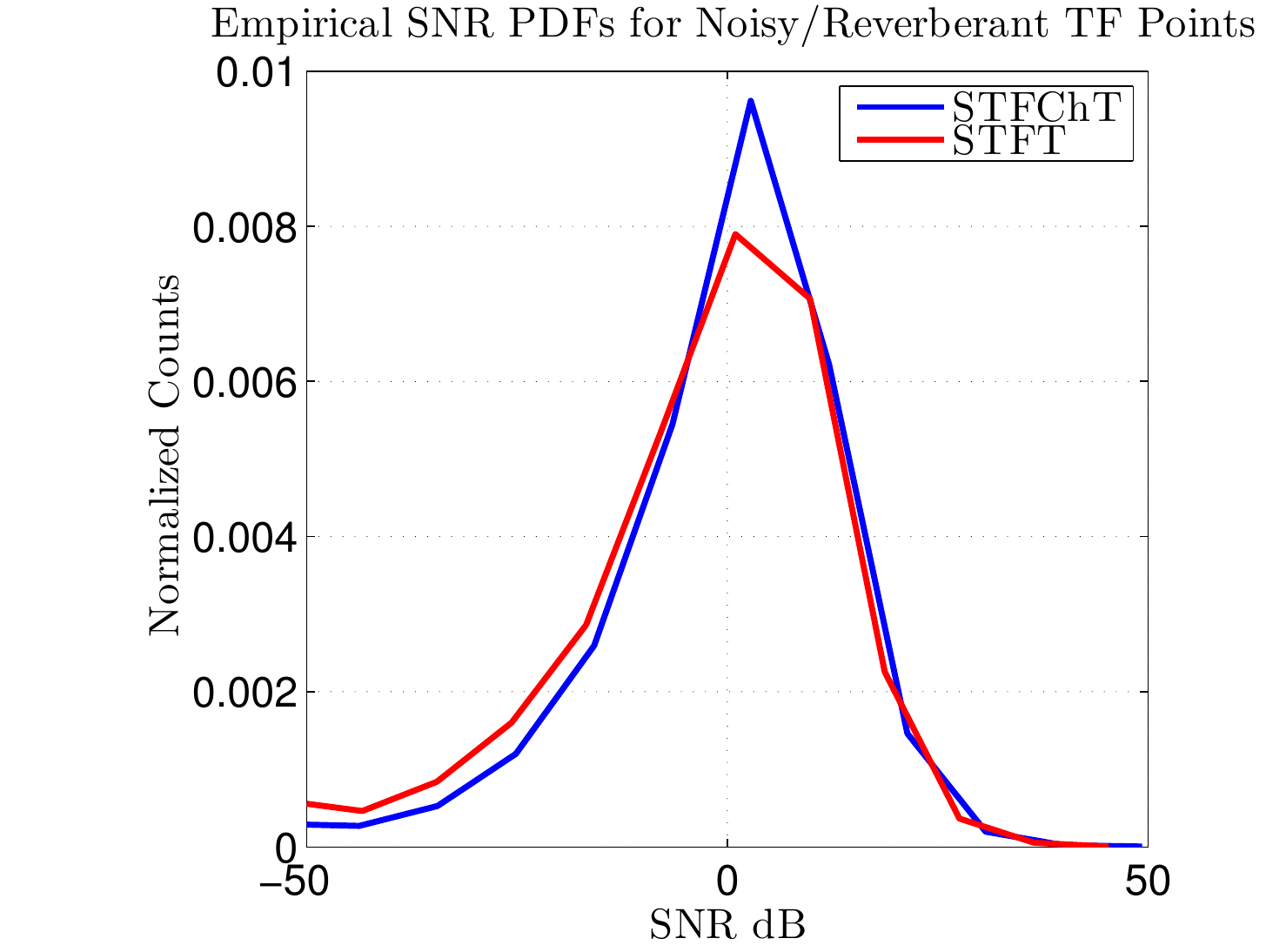}
	\caption{{\bf Empirical distribution of local SNR values in time-frequency points containing only noise and reverberation.} Given for the synthetic chirp signals in figure \ref{fig:snrmap}. Notice that the STFT and STFChT have similar distributions for SNR in these signal-free time-frequency bins.} \label{fig:histreverb}
\end{minipage}
\end{figure}

Figure \ref{fig:snrmap} shows these oracle SNR values for the STFT and the STFChT representations of the chirps. Figures \ref{fig:histdir} and \ref{fig:histreverb} show empirical probability density functions (PDFs) for the SNR values under two cases: time-frequency bins containing direct signal, noise and reverberation, and bins containing only noise and reverberation. We designate direct bins as the ones in which the direct signal should ideally fall given our knowledge of the synthetic test signals, and the noisy/reverberant bins make up the remainder. 

As can be seen in the two plots in Figure \ref{fig:snrmap}, the STFChT (right) appears to better lock on to the harmonics despite the noise and reverberation, whereas the STFT (left) smears out the energy in time and frequency. Figure \ref{fig:histdir} shows that the expected SNR in STFChT bins is higher than the expected SNR in STFT bins, while figure \ref{fig:histreverb} shows that the distribution of the SNR in noisy and reverberant bins is unchanged from STFT to STFChT. The STFChT effectively partitions more direct signal power from noise and reverbation.
Since HMMSE-LSA applies gains to individual time-frequency bins, the more the STFChT can partition direct signal power from noise and reverbation, the better performance will be. Thus, when a noise and reverberation dominated time-frequency bin is suppressed, less speech power is lost, and fewer speech artifacts are created. 


Moreover, concentrating direct-path signal power prevents HMMSE-LSA from over-suppressing the speech signals, which is a common problem when enhancing in the STFT domain. Capp\'{e} analyzed \cite{cappe_elimination_1994} how the original Ephraim and Malah LSA estimator \cite{ephraim_speech_1984} tends to greatly reduce musical noise artifacts. Musical noise artifacts are an unnatural disturbance in speech enhanced using MMSE-LSA, and is caused by enhanced noise-only bands having spectral peaks that sound like random narrowband tones \cite{cappe_elimination_1994}. MMSE-LSA tends to have less artifacts than Wiener filtering or spectral subtraction.

Capp\'{e} observed that a high {\it a posteriori} SIR $\gamma(d,k)$ causes more attenuation compared to a standard Wiener gain, especially when the {\it a priori} SIR $\xi(d,k)$ is small; $\gamma(d,k)$ provides a ``correction factor" when the $\xi(d,k)$ has been incorrectly estimated.

Considering this observation, Capp\'{e} described two cases:
\begin{enumerate}
\item $\gamma(d,k) \leq 0$dB, i.e. noise-dominated time-frequency bins: in this case, $\xi(d,k)$ is a highly smoothed version of  $\gamma(d,k)$. This smoothing eliminates spectral peaks in noise-only regions
\item $\gamma(d,k) > 0$dB, i.e. speech-dominated time-frequency bins: in this case, $\xi(d,k)$ tends to follow $\gamma(d,k)$ with a one-frame delay.
\end{enumerate}
We have seen that the STFChT of a harmonic signal concentrates more direct signal energy into only a few bins as compared to the STFT. Thus, according to point 2 above, when only a few bins correspond to speech, in these bins the {\it a priori} SIR $\xi(d,k)$ will closely follow $\gamma(d,k)$. Furthermore, since the SNR distribution in noise and reverberation-dominated bins is similar between the STFT and STFChT, the advantageous smoothing mentioned in point 1 will reduce spectral peaks and hence tonal artifacts.

\begin{figure}[h]
\includegraphics[width=0.95\linewidth]{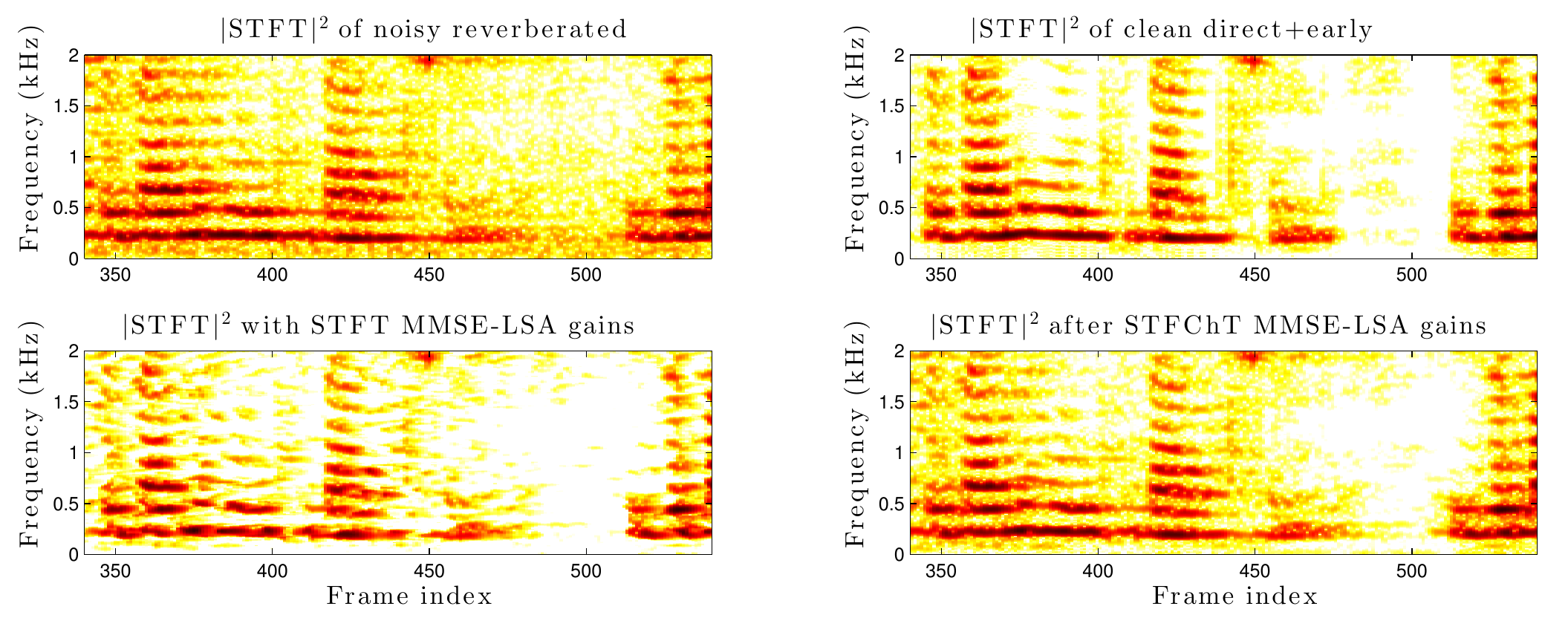}
\caption{{\bf Spectrogram comparisons of STFT-based HMMSE-LSA to STFChT-based HMMSE-LSA}. Upper left: noisy audio. Upper right: ideal clean signal with some early reflections, which is ground truth to be recovered. Lower left: spectrogram of enhancement using STFT-based HMMSE-LSA. Lower right: spectrogram of enhancement using STFChT-based HMMSE-LSA. The comparison between lower left and lower right shows that STFChT exhibits less over-suppresion of speech energy.} \label{fig:sgram_compare}
\end{figure}

An example of the STFChT providing less over-suppression is shown in figure \ref{fig:sgram_compare}. The figure shows a clip of a noisy, reverberated speech signal (upper left panel) using the same RIR and noise as used for the synthetic chirps in figures \ref{fig:snrmap} through \ref{fig:histreverb}. The upper right panel shows the direct signal plus early reflections that are desired to be recovered. STFT-based HMMSE-LSA processing exhibits over-suppression of direct speech energy (lower left), while the STFChT better preserves the direct speech signal (lower right).

\subsection{Iterative enhancement and parameter tuning}

Our enhancement method can be improved by subsequent iterations. Iterative enhancement proceeds by successively running our above algorithm multiple times on a noisy utterance and taking a weighted convex combination of these outputs. In general, the output of iterative enhancement is
\begin{equation}
\hat{x}_{iter}[n]=\sum_{i=1}^I a_i\hat{x}_{i\mathrm{x}}[n]
\end{equation}
where $\hat{x}_{i\mathrm{x}}[n]$ is the noisy single-channel audio $y[n]$ processed $i$ times by an enhancement algorithm, $I$ is the maximum number of iterations, and $\{a_i\}_{1:I}$ are convex mixing weights (that is, the $a_i$ are nonnegative and $\sum_{i=1}^I a_i = 1$). In particular, we found that performance was best improved using a convex combination of once- and twice-iterated processing; thus, we set $I=2$. The second iteration of processing uses reverberation parameters estimated during the first iteration of processing (e.g., $T_{60}$ time). Iterative processing is done on single-channel data, and can serve as a post-filter for a beamformer.

We performed experiments to tune the parameters of iterative enhancement. Our goal was not only to discover the optimal iterative mixing parameter $a$, but to also choose the best analysis window duration $T_{win}$. For $I=2$, the convex weights are parameterized by $a$, with $a_1=a$ and $a_2=1-a$, and $0\leq a \leq 1$. The degree of iterative enhancement is given by $(1-a)$, since a larger $(1-a)$ indicates more of the twice-processed audio in the output. To tune these parameters, we choose $30$ random utterances from each of the $6$ SimData conditions, which are all permutations of the three rooms (room1, room2, and room3) and two distances (near and far). We tried both STFT- and STFChT-based processing on these utterances.

\begin{figure}[h]
  \centering
\includegraphics[width=\linewidth]{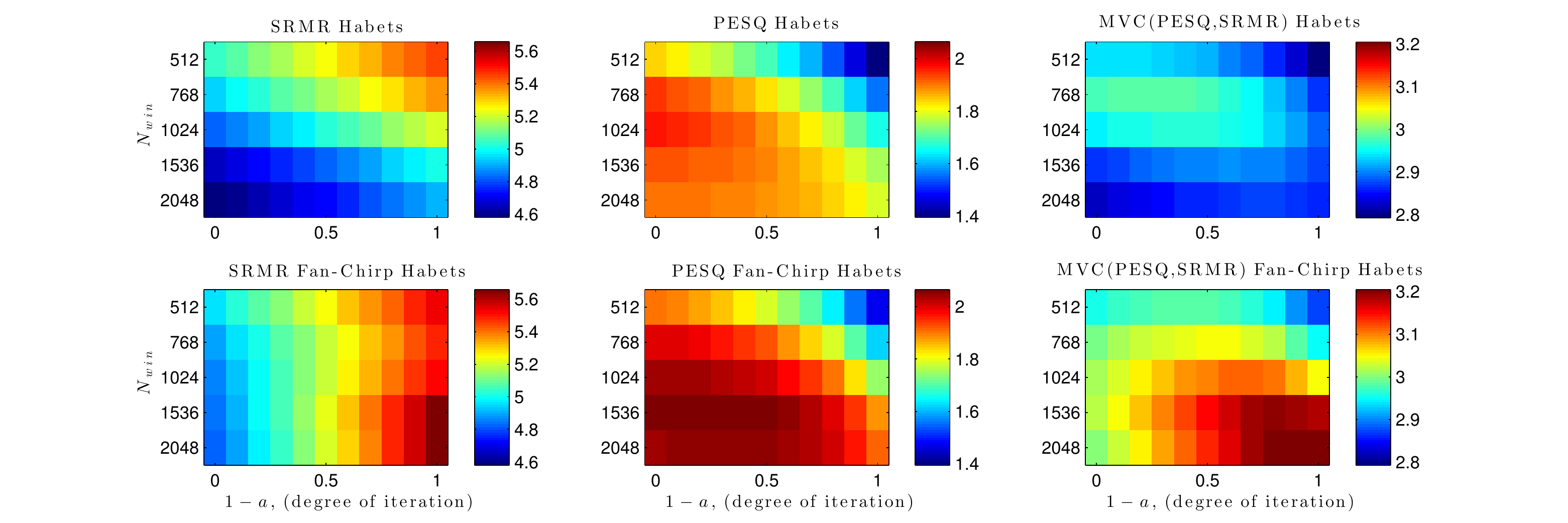}
  \caption{{\bf Performance of STFT-based and STFChT-based HMMSE-LSA versus degree of iteration and window length on development SimData.} Plots illustrating the trade-off between speech quality (measured by PESQ) and dereverberation (measured by SRMR). In general, the STFChT-based method achieves superior speech quality and dereverberation.
      } \label{fig:results_iterated_SRMR_PESQ}
\end{figure}

Figure \ref{fig:results_iterated_SRMR_PESQ} shows the PESQ and SRMR scores versus $T_{win}$ and $(1-a)$. The results reveal an interesting trade-off between speech quality (measured by PESQ) and dereverberation (measured by SRMR): a higher degree of iteration results in more dereverberation, at the cost of speech quality. These results also demonstrate the ability of the STFChT to increase analysis window duration. For STFT processing, a window duration of $64$ ms is optimal, while for STFChT processing, a window duration of $96$ or $128$ ms is optimal.

To discover the optimal trade-off between speech quality and dereverberation, we perform a minimum variance combination (MVC) of the PESQ and SRMR scores. This combination is given by
\begin{equation}
C=(1-\hat{c})\cdot\mathrm{SRMR}+\hat{c}\cdot\mathrm{PESQ}
\end{equation}
where
\begin{equation}
\begin{aligned}
\hat{c}= \underset{c}{\text{argmin}}
&  \sum_{i} \left[ (1-c)\cdot\mathrm{SRMR}_i+c\cdot\mathrm{PESQ}_i\right ]^2
\end{aligned}
\end{equation}
where $i$ runs over the indices of all combinations of $T_{win}$ and $(1-a)$ that are being tested. This produces the minimum variance combination of PESQ and SRMR, which takes into account the correlation between the two measures and their variances.

For STFT-based HMMSE-LSA (top panels), shorter windows ($T_{win}=48$ ms or $64$ ms) tend to give the best PESQ/SRMR values, while for STFChT-based HMMSE-LSA (bottom panels), longer windows ($T_{win}=96$ ms or $128$ ms) tend to give better results. In general, a higher degree of iteration ($(1-a)=1$) provide better suppression of reverberation, at the expense of speech quality. An iteration degree of $(1-a)=0.3$ yields the best PESQ score. An optimal trade-off between PESQ and SRMR, as measured by the MVC between them, is $N_{win}=96$ ms and $(1-a)=0.7$ (lower right). Overall, STFChT Habets achieves higher objective scores on both PESQ and SRMR.

Using the information above, we reprocessed the REVERB SimData using a window duration of $96$ ms, and degrees of iteration of $(1-a)=0.3$ and $(1-a)=0.7$. A degree of iteration of $(1-a)=0.3$ performed best out of these two (a degree of iteration of $(1-a)=0.7$ gave worse objective metrics, except for SRMR). These best scores are shown in tables \ref{fig:resultsSimData_eval} and \ref{fig:resultsRealData_eval}.

\section{Implementation}

Our algorithms are implemented in MATLAB, and we use utterance-based processing. The algorithm starts by using the utterance data to estimate the $T_{60}$ time of the room using the blind algorithm proposed by L\"{o}llmann et al. \cite{lollmann_improved_2010}. Multichannel utterance input data is concatenated into a long vector, and as recommended by L\"{o}llmann et al., noise reduction is performed beforehand. We use Loizou's implementation \cite{loizou_speech_2007} of Ephraim and Malah's LSA \cite{ephraim_speech_1985} for this pre-enhancement.

\begin{figure}[h!]
  \centering
\includegraphics[width=0.75\linewidth]{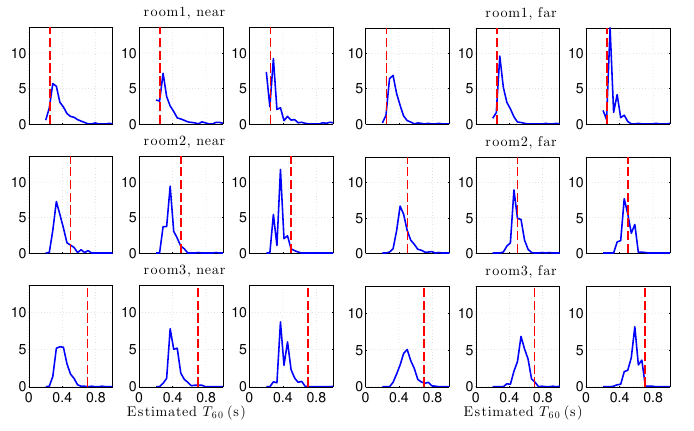}
  \caption{{\bf Performance of blind $T_{60}$ estimation algorithm}
      Sample probability density functions of estimated $T_{60}$ time measured on SimData evaluation dataset (these results were not used to tune the algorithm). For each condition, left plot is for 1-channel data, center plot is for 2-channel data, and right plot is for 8-channel data.} \label{fig:T60}
      \end{figure}

Figure \ref{fig:T60} shows empirically-estimated probability density functions (PDFs) of the $T_{60}$ estimation performance using this approach. These plots show that $T_{60}$ estimation \cite{lollmann_improved_2010} precision generally improved with increasing amounts of data (i.e., with more channels), although for some conditions $T_{60}$ estimates were inaccurate. Vertical dashed lines indicate approximate $T_{60}$ times given by REVERB organizers \cite{kinoshita_reverb_2013}.

\subsection{Spatial processing for multichannel data}

For multichannel data, we estimate the direction of arrival (DOA) by cross-correlating oversampled data between channels. That is, we compute a $N_{ch}$-length vector of time delays ${\bf d}$ with $d_1=0$ and $d_i$, $i$=2,...,$N_{ch}$ given by
\begin{equation}
d_i = \underset{k}{\mathrm{argmax}} \frac{r_{1i}[k]}{Uf_s},
\end{equation}
where $r_{1i}[k]=\sum_n x_1[n]x_i[n-k]$, $U$ is the oversampling factor, and $c=340$ meters per second, the approximate speed of sound in air.

Given a time delay vector ${\bf d}$, the DOA estimate is given by the solution to ${\bf P}\hat{\bf a} = \frac{1}{c}{\bf d}$, 
where $\hat{\bf a}$ is a $3\times 1$ unit vector representing the estimated DOA of the speech signal and ${\bf P}$ is a $N_{ch} \times 3$ matrix containing the Cartesian $(x,y,z)$ coordinates of the array elements. For example, for an eight-element uniform circular array, $P_{i1}=x_i=r \cos(i\pi / 4)$, $P_{i2}=y_i=r \sin(i\pi / 4)$, and $P_{i3}=z_i=0$ for $i=0,1,...,7$, where $r$ is the array radius.

\begin{figure}[h]
\centering
\includegraphics[width=0.75\linewidth]{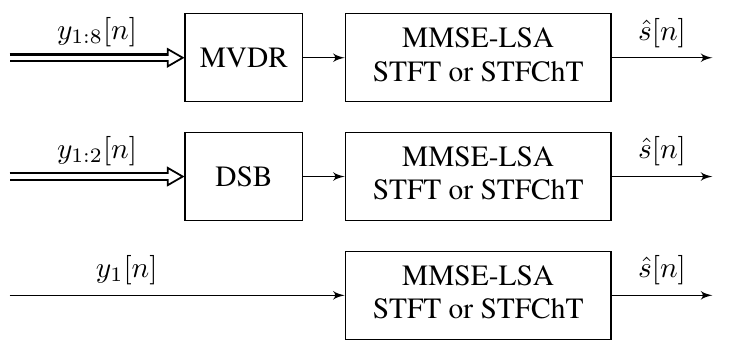}
\caption{{\bf Block diagrams of processing} For 8-channel data using a minimum variance distortionless response (MVDR) beamformer (top), 2-channel data using a delay-and-sum beamformer (DSB, middle), and 1-channel data (bottom).} \label{fig:blk_diag}
\end{figure}

For the 8-channel case, the estimated DOA is used to form the steering vector ${\bf v}^H(f)$ for a frequency-domain minimum variance distortionless response (MVDR) beamformer applied to the multichannel signal. The weights ${\bf w}^H(d,f)$ for the MVDR are \cite[equations (6.14-15)]{van_trees_optimum_2002}
\begin{equation}
{\bf w}^H(d,f) = \frac{ {\bf v}^H(f) {\bf S}_{{\bf y}{\bf y}}^{-1}(d,f) }{{\bf v}^H(d,f) {\bf S}_{{\bf y}{\bf y}}^{-1}(d,f) {\bf v}(d,f)},
\end{equation}
where ${\bf S}_{{\bf y}{\bf y}}(d,f)$ is the spatial covariance matrix at frequency $f$ and frame $d$ estimated using $N$ snapshots $Y(d-n,f)$ for $-N/2 \leq n < N/2$ and ${\bf v}$ is given by
\begin{equation}
{\bf v}(f) = \exp\left(j \frac{2\pi f}{c} {\bf P} \hat{\bf a} \right).
\end{equation}
Our MVDR implementation uses a 512-sample long Hamming window with 25\% overlap, a 512-point FFT, and $N=24$ snapshots for spatial covariance estimates. For 2-channel data, we use a delay-and-sum beamformer to enhance the signal with the delay given by the DOA estimate. Single-channel data is enhanced directly by the single-channel HMMSE-LSA algorithm. A block diagram of these three cases is shown in figure \ref{fig:blk_diag}.

\begin{table*}[t]
\caption{Summary of speech enhancement and ASR results on single- and eight-channel REVERB evaluation data (SimData/RealData, RealData results given when applicable). Arrows indicate whether a higher or lower metric is better.} \label{table:summary_eval1ch}
      \begin{tabular}{c | cccccc}
        \begin{tabular}{@{}c@{}} Beamforming, \\ TF type, \\ Window duration \end{tabular}
        							& \begin{tabular}{@{}c@{}} Mean$|$Med. \\ CD [dB] \\ ($\downarrow$)  \end{tabular}
        							& \begin{tabular}{@{}c@{}} SRMR \\  ($\uparrow$) \end{tabular}
        							& \begin{tabular}{@{}c@{}} Mean$|$Med. \\ LLR \\ ($\downarrow$)  \end{tabular}
        							& \begin{tabular}{@{}c@{}} Mean$|$Med. \\ FWSegSNR \\ $[$dB$]$ ($\uparrow$) \end{tabular}  
        							& \begin{tabular}{@{}c@{}} PESQ \\ ($\uparrow$) \end{tabular}
        							& \begin{tabular}{@{}c@{}} WER \\ $[$\%$]$ \\ ($\downarrow$) \end{tabular}
        						\\ \hline \\[1.25ex]
        No enh.
                  & 3.97$|$3.68 & 3.68/3.18 & 0.57$|$0.51 & 3.62$|$5.39  & 1.48 & 11.97/30.27 \\[2.5ex] \hline \hline \\
        \begin{tabular}{@{}c@{}} None, \\ STFT, \\ $32$ ms  \end{tabular}
                  & 3.87$|$3.48  & {\bf 4.79}/{\bf 5.80} & 0.68$|$0.58 & 6.72$|$7.62 & 1.53 & 12.32/33.37 \\[4ex] \hline \\
        \begin{tabular}{@{}c@{}} None, \\ STFT, \\ $128$ ms  \end{tabular}
                  & 3.84$|$3.51 & 4.28/4.21 & {\bf 0.54}$|${\bf 0.47} & 4.65$|$6.71 & 1.59 & {\bf 10.20}/{\bf 28.23} \\[4ex] \hline \\
        \begin{tabular}{@{}c@{}} None, \\ STFChT, \\ $128$ ms  \end{tabular}
                  & {\bf 3.57$|$3.07} & 4.55/4.85 & 0.57$|$0.49 & {\bf 7.07$|$8.60 } & {\bf 1.69 } & { 11.21}/{32.03} \\[4ex] \hline \hline \\
\begin{tabular}{@{}c@{}} 8ch MVDR, \\ No enh. \\ \end{tabular}
                  & 3.15$|$2.81 & 3.96/4.03 & 0.44$|$0.38 & 5.95$|$8.45 & 1.80 & 8.82/21.68 \\[4ex] \hline \\
\begin{tabular}{@{}c@{}} 8ch MVDR, \\ STFT, \\ $32$ ms  \end{tabular}
                  & 3.56$|$3.23 & 4.77/{\bf 6.90} & 0.61$|$0.50 & 8.06$|$8.47 & 1.83 & 9.84/32.19 \\[4ex] \hline \\
         \begin{tabular}{@{}c@{}} 8ch MVDR, \\ STFT, \\ $128$ ms  \end{tabular}
                  & 3.18$|$2.83 & 4.56/5.31 & 0.43$|$0.38 & 6.79$|$9.31 & 1.94 & {\bf 7.62}/{\bf 19.84} \\[4ex] \hline \\
         \begin{tabular}{@{}c@{}} 8ch MVDR, \\ STFChT, \\ $128$ ms  \end{tabular}
                  & {\bf 2.97$|$2.49} & {4.82}/6.33 & {\bf 0.43$|$0.37} &  9.21$|${\bf 10.63} & {2.10}  & {8.18}/{22.21}  \\[4ex] \hline \\
         \begin{tabular}{@{}c@{}} 8ch MVDR, \\ Iterated STFChT, \\ $(1-a)$=0.3, $96$ ms  \end{tabular}
                  & {3.33 $|$ 2.78} & {\bf 5.03}/6.78 & {0.44$|$0.38} & {\bf 9.37}$|$10.54 & {\bf 2.14}  & { 8.18}/{22.21}  \\[4ex] \hline \\
      \end{tabular}
\end{table*}

\subsection{Time-frequency analysis-synthesis}

We tried three analysis-synthesis domains for the HMMSE-LSA enhancement algorithm: the STFT with a short window, the STFT with a long window, and the STFChT. The STFT with a short window uses 512-sample long ($T=32$ms) Hamming windows, a frame hop of 128 samples, and an FFT length of 512. Short-window STFT processing is chosen to match conventional speech processing window lengths. The STFT with a long window uses 2048-sample long ($T=128$ms) Hamming windows, a frame hop of 128 samples, and an FFT length of 3262. Long-window STFT processing is intended to match the parameters of STFChT processing for a direct comparison. STFChT processing uses an analysis duration of 2048 samples, a Hamming analysis window, a frame hop of 128 samples, an FFT length of 3262, oversampling factor of 8, and a set of possible analysis chirp rates $\mathcal{A}$ consisting of 21 equally spaced $\alpha$s from -4 to 4.

The forward STFChT, given by (\ref{eq:stfcht}), proceeds frame-by-frame, estimating the optimal analysis chirp rate $\hat{\alpha}_d$ using (\ref{eq:glogs}), oversampling in time, warping, applying an analysis window, and taking the FFT. Then HMMSE-LSA weights are estimated frame-by-frame and applied in the STFChT domain, and the enhanced speech signal is reconstructed using the inverse STFChT.

For all methods, noise estimation is performed with a decision-directed method and simple online updating of the noise variance. Voice activity detection to determine if a frame is noise-only or speech-plus-noise is done using Loizou's method, which compares the following quantity to a threshold $\eta_\mathrm{thresh}$:
\begin{equation} \label{eq:vad_stat}
\eta(d) = \sum_k \ln \gamma(d,k) \frac{\xi(d,k)}{1+\xi(d,k)} - \ln (1+\xi(d,k)).
\end{equation}
If $\eta(d)<\eta_\mathrm{thresh}$, the frame is determined to be noise-only and the noise variance is updated as $\lambda_v(d,k) = \mu_{v} \lambda_v(d-1,k) + (1-\mu_{v}) |Y(d,k)|^2$, with $\mu_v=0.98$ and $\eta_\mathrm{thresh}=0.15$.

For our implementation of Habets's joint dereverberation and noise reduction algorithm, we used Loizou's implementation \cite{loizou_speech_2007} of Ephraim and Malah's LSA \texttt{logmmse} MATLAB algorithm as a foundation. The forward STFChT code was written by Cancela et al. \cite{cancela_fan_2010}. 
We wrote our own MATLAB implementation of the inverse STFChT.

For 8-channel data, the MVDR and the STFChT require the most computation. For 1-channel and 2-channel data, the STFChT requires the most computation. For the STFChT, most of the computation is used to compute the GLogS for estimation of the analysis chirp rate $\hat{\alpha_d}$ (\ref{eq:glogs}) for each frame. Note that this computation could be easily parallelized in hardware.


\section{Experiments}

We compare the effectiveness of using the STFT or the STFChT as the analysis-synthesis domain for HMMSE-LSA algorithm described in section \ref{ssec:mmse_lsa}. The tasks are the two tracks of the REVERB challenge: speech enhancement and automatic speech recognition.

We evaluate our algorithms on the REVERB challenge dataset \cite{kinoshita_reverb_2013}. The data consists of both simulated and real reverberated speech. Simulated data (SimData) are created by convolving utterances from the Wall Street Journal Cambridge read news (WSJCAM0) corpus \cite{robinson_wsjcam0:_1995} with measured room impulse responses for three different reverberant rooms and at two distances: a near distance of about 0.5 meters and a far distance of about 2 meters. Recorded air conditioning noise is added at about 20dB signal-to-noise ratio (SNR). Real data (RealData) are actual recordings of male and female speakers from the multichannel Wall Street Journal audio-visual (MC-WSJ-AV) corpus \cite{lincoln_multi-channel_2005} reading prompts in a noisy (air conditioning noise at about 20dB SNR) and reverberant room, at two distances: a near distance of 1 meter and a far distance of 2.5 meters. 

A summary table of our results is shown in table \ref{table:summary_eval1ch} for single- and eight-channel data. For single-channel data, the top part of table \ref{table:summary_eval1ch} shows that STFChT processing yields superior enhancement results, but long-window ($T_{win}=128$ ms) STFT processing yields superior recognition results. In the bottom part of table \ref{table:summary_eval1ch}, results for eight-channel data indicate that performing multichannel STFChT processing generally yields superior enhancement as compared to STFT processing. For recognition, STFT processing with a long window achieves the lowest WERs.

\subsection{Speech Enhancement Results}


We score the enhanced audio using the same metrics used for the REVERB challenge, which includes segmental frequency-weighted SNR (FWSegSNR), cepstral distance (CD), source-to-reverberation modulation ratio (SRMR) \cite{falk_non-intrusive_2010}, log likelihood ratio (LLR), and perceptual evaluation of speech quality (PESQ) \cite{rix_perceptual_2001}. All of these metrics are intrusive (meaning that they required clean reference signals) except for SRMR, which is the only non-intrusive metric. Since RealData does not have clean reference signals, SRMR is the only metric that can be run on RealData. Note that the precision of the scores reported is possibly lower than the precision implied by the number of significant digits reported. For consistency with the work of others, we chose to have our table entries match the precision used by the REVERB challenge results\footnote{\texttt{reverb2014.dereverberation.com/result\_se.html}}.

\begin{figure}[h]
\includegraphics[width=\linewidth]{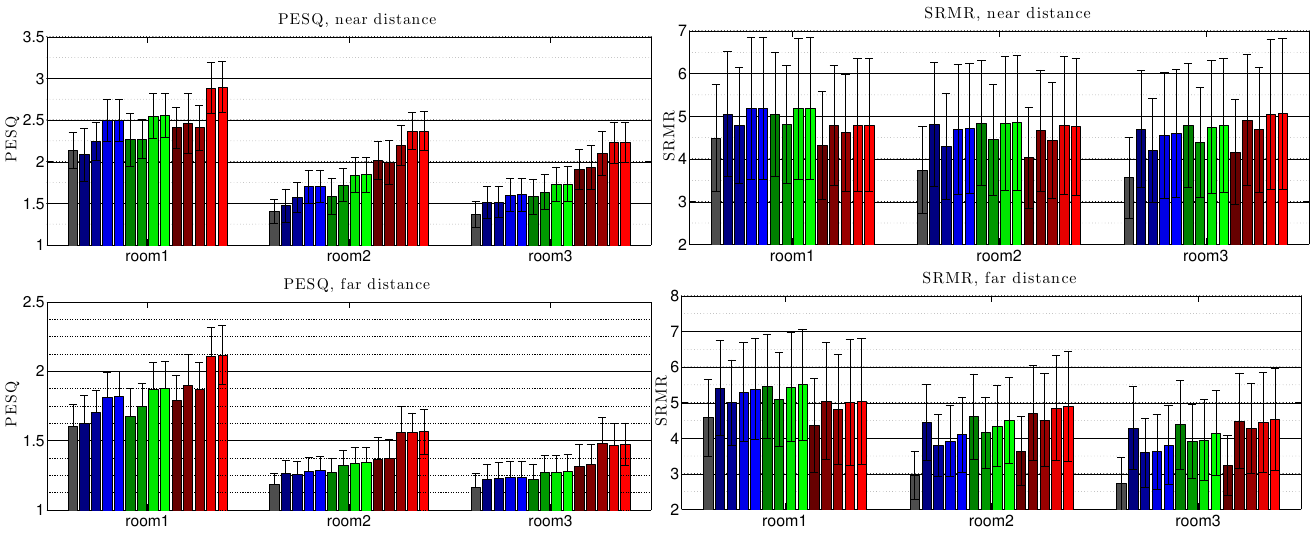} \\
\centering
\includegraphics[width=0.225\linewidth]{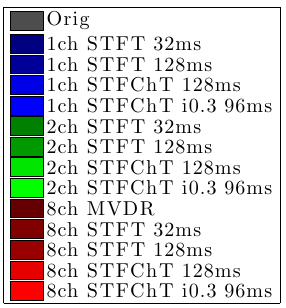}
\caption{{\bf PESQ and SRMR results for SimData evaluation set} Upper plots are near distance condition, lower plots are far distance condition. ``i0.3'' indicates iterative enhancement with $(1-a)=0.3$.}
\label{fig:pesq_bar}
\end{figure}
\begin{figure}[h]
\centering
\includegraphics[width=0.5\linewidth]{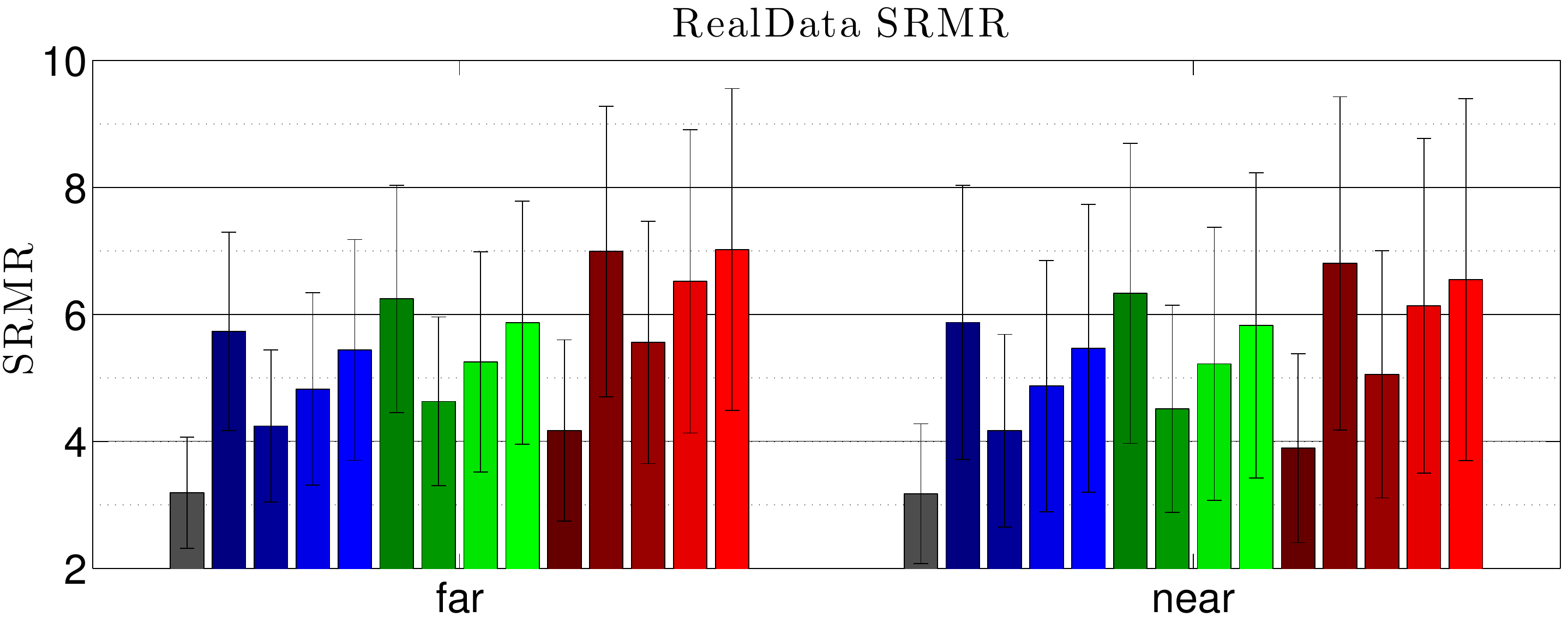}
\caption{{\bf SRMR results for RealData evaluation set} Same legend as figure \ref{fig:pesq_bar}.} \label{fig:resultsRealData_eval_bar}
\end{figure}

Our results on REVERB evaluation data are shown in figures \ref{fig:pesq_bar} and \ref{fig:resultsRealData_eval_bar} and tables \ref{fig:resultsSimData_eval} and \ref{fig:resultsRealData_eval}. Tables \ref{fig:resultsSimData_eval} and \ref{fig:resultsRealData_eval} also include computation times in terms of the real-time factor (RTF), which we define as total processing time divided by total data time. We choose to display PESQ (Perceptual Evaluation of Speech Quality) \cite{rix_perceptual_2001} and SRMR (source-to-reverberation modulation energy ratio) \cite{falk_non-intrusive_2010} more prominently because the former is the ITU-T standard for voice quality testing \cite{itu-t_p.862.2_wideband_2007} and the latter is both a measure of dereverberation and the only non-intrusive measure that can be run on RealData (for which the clean speech is not available).

For SimData, STFChT-based enhancement always performs better in terms of PESQ than STFT-based enhancement using either a short (512-sample) window or a long (2048-sample) window, for the 8-, 2-, and 1-channel cases (except for 8-channel, far-distance data in room 3). Informal listening tests revealed an oversuppression of speech and some musical noise artifacts in STFT processing, while STFChT processing did not exhibit oversuppression or musical noise artifacts. The oversuppression of direct-path speech by STFT processing can be seen in the spectrogram comparisons shown in figure \ref{fig:sgram_compare}. In terms of SRMR, STFChT processing yields equivalent or slightly worse SRMR scores than long-window STFT processing for the 8-, 2-, and 1-channel cases (except for 8-channel, near-distance data, where STFChT processing does slightly better).  Informal listening indicated that although STFT processing reduced reverberation more, it came at the cost of oversuppression of speech.
One issue with these SRMR comparisons, however, is that the variance of the SRMR scores is quite high. Thus, for SimData, STFChT processing achieves better perceptual audio quality while still achieving almost equivalent dereverberation compared to STFT processing.

\subsection{Automatic Speech Recognition Results}

For ASR experiments, we use the GMM-HMM recognizer implemented in Kaldi\footnote{\texttt{www.mmk.ei.tum.de/\textasciitilde wen/REVERB\_2014/kaldi\_baseline.tar.gz}} by Weninger et al. \cite{weninger_merl/melco/tum_2014}. The front-end of the ASR concatenates nine adjacent frames of 13 Mel-frequency cepstral coefficients (MFCCs) each and uses linear discriminant analysis (LDA) and semi-tied covariance (STC) \cite{gales_semi-tied_1999} to reduce these features down to 40 dimensions. The recognizer includes per-utterance feature-based maximum likelihood linear regression (fMLLR) for adaptation and uses minimum Bayes risk (MBR) for decoding. Optional discriminative training is performed using boosted maximum mutual information (bMMI). Tuning the language model weight and beam-width further optimizes the decoding.

We use HMMSE-LSA in the STFT and STFChT domains to enhance reverberant and noisy data before feeding the enhanced audio to the recognizer. Unlike Weninger et al., we found that using noisy multicondition training data with enhanced audio could improve WER versus using noisy multicondition training data with noisy audio. However, the lowest WERs occurred when the recognizer was trained with pre-enhanced noisy multicondition data (pre-enhanced with the single-channel part of the corresponding enhancement algorithm) and run on enhanced audio.

To show the effect of various recognizer optimizations, recognition results are shown in tables \ref{table:asr1} and \ref{table:asr2},  We show two decimal places to be consistent with REVERB challenge results\footnote{\texttt{reverb2014.dereverberation.com/result\_asr.html}}. For both development and evaluation data, HMMSE-LSA with a long-window STFT ($T_{win}=128$ ms) performed best for both 8-channel and single-channel data.

It is interesting that STFT-based enhancement yields better ASR performance over STFChT-based enhancement, especially since STFChT-based enhancement achieves better objective enhancement scores. We hypothesize that the better ASR performance using STFT-based enhancement results from the STFChT adding distortions to vocal tract dynamics. Though the STFChT concentrates harmonic signal energy for voiced speech, which results in better enhancement as discussed in section \ref{ssec:advantage}, this concentration of energy comes with the trade-off of distortion to the spectral envelope of the windowed frame, with distortions increasing with increasing chirp rates. Such distortions of the spectral envelopes result in less discriminative ASR features, thus increasing phone error rate, and in turn word error rate.

\section{Conclusion}

In this paper, we have demonstrated the advantages of a new transform domain for speech enhancement: the short-time fan-chirp transform (STFChT). By estimating linear fits in the instantaneous fundamental frequency of voiced speech signals, the STFChT is more coherent with speech signals over longer durations, which allows extension of analysis window duration. In turn, this increased window duration concentrates more direct-path signal into time-frequency bins, which enables superior enhancement results in terms of objective metrics like PESQ and SRMR. We also performed ASR experiments on both STFT- and STFChT-based enhancement. Interestingly, despite better objective enhancement scores, we observed that long-window (128 ms) STFT processing yielded the lowest WERs.

The utility of the STFChT warrants further investigation. Interesting future directions include moving beyond linear models of instantaneous frequency. Combinations of the STFChT and other coherence-extending transforms with deep neural network (DNN) enhancement and recognition methods could yield further performance improvements.

\section*{Acknowledgments}
   We wish to thank Derek Huang for his help with the Kaldi tools. This work is funded by ONR contract N00014-12-G-0078, delivery order 0013, and ARO grant number W911NF1210277.

\begin{table*}%
\caption{Results for SimData evaluation set.} \label{fig:resultsSimData_eval}
\centering
\begin{minipage}{\linewidth}
\centering
\scalebox{1}{\bf SimData summary} \\
\resizebox{\chartWidthScale\linewidth}{!}{%
\hspace{-90pt}
\begin{tabular}{||y{16pt} y{73pt} y{73pt} y{73pt} y{43pt} y{43pt} y{43pt} y{43pt} y{43pt} y{53pt} y{53pt} ||}\hline
 Ch. & Method & Comp. time (RTF) & Mean CD & Median CD & SRMR & Mean LLR & Median LLR & Mean FWSegSNR & Median FWSegSNR & PESQ \\
 &Orig & --- & {\it 3.97} &{\it 3.68} &{\it 3.68} &{\it 0.57} &{\it 0.51} &{\it 3.62} &{\it 5.39} &{\it 1.48}\\ \hline
8 &STFT 32ms/128ms & 2.59 / 2.65 &3.56 / 3.18 &3.23 / 2.83 &4.77 / 4.56 &0.61 / {\bf 0.43} &0.50 / 0.38 &8.06 / 6.79 &8.47 / 9.31 &1.83 / 1.94\\
8 &STFChT 128ms & 5.97 &{\bf 2.97} &{\bf 2.49} &4.82 &{\bf 0.43} &{\bf 0.37} &9.21 &{\bf 10.63} &2.10\\
8 &STFChT i0.3 96ms & 8.56 &3.06 &2.57 &{\bf 5.03} &0.44 &0.38 &9.37 &10.54 &{\bf 2.14}\\
2 &STFT 32ms/128ms & 0.68 / 0.70 &3.80 / 3.57 &3.42 / 3.22 &4.86 / 4.47 &0.65 / {\bf 0.49} &0.55 / {\bf 0.44} &7.26 / 5.46 &7.93 / 7.86 &1.60 / 1.66\\
2 &STFChT 128ms & 2.87 &{\bf 3.33} &{\bf 2.83} &4.75 &0.51 &0.45 &7.68 &9.19 &1.77\\
2 &STFChT i0.3 96ms & 5.47 &3.37 &2.84 &{\bf 5.04} &0.51 &{\bf 0.44} &{\bf 8.06} &{\bf 9.32} &{\bf 1.81}\\
1 &STFT 32ms/128ms & 0.35 / 0.37 &3.87 / 3.84 &3.48 / 3.51 &4.79 / 4.28 &0.68 / {\bf 0.54} &0.58 / {\bf 0.47} &6.72 / 4.65 &7.62 / 6.71 &1.53 / 1.59\\
1 &STFChT 128ms & 2.60 &{\bf 3.57} &3.07 &4.55 &0.57 &0.49 &7.07 &8.60 &1.69\\
1 &STFChT i0.3 96ms & 5.19 &3.59 &{\bf 3.06} &{\bf 4.83} &0.57 &0.49 &{\bf 7.57} &{\bf 8.89} &{\bf 1.72}\\
\hline
\end{tabular}

}
\end{minipage} \\
\end{table*}

\begin{table*}%
\caption{Results for RealData evaluation set.} \label{fig:resultsRealData_eval}
\centering
\begin{minipage}{0.5\linewidth}
\centering
\scalebox{1}{\bf RealData summary} \\
\resizebox{\linewidth}{!}{%
\begin{tabular}{|y{16pt} y{73pt} y{73pt} y{73pt}|}\hline
 Ch. & Method & Comp. time (RTF) & SRMR \\
 &Orig & --- &{\it 3.18}\\ \hline
8 &STFT 32ms/128ms & 2.54 / 2.60 &{\bf 6.90} / 5.31\\
8 &STFChT 128ms & 4.32 &6.33\\
8 &STFChT i0.3 96ms & 6.59 &6.78\\
2 &STFT 32ms/128ms & 0.70 / 0.77 &{\bf 6.29} / 4.57\\
2 &STFChT 128ms & 2.51 &5.24\\
2 &STFChT i0.3 96ms & 4.78 &5.85\\
1 &STFT 32ms/128ms & 0.50 / 0.56 &{\bf 5.80} / 4.21\\
1 &STFChT 128ms & 2.27 &4.85\\
1 &STFChT i0.3 96ms & 4.54 &5.45\\
\hline
\end{tabular}

}
\end{minipage}
\end{table*}

\begin{table*}
\caption{ASR results for REVERB development set using the Kaldi baseline recognizer by Weninger et al. \cite{weninger_merl/melco/tum_2014}. Results are word error rates (WERs) in \% for SimData/RealData. Beamforming describes the spatial processing used, time-frequency (TF) type describes the analysis-synthesis domain for Habets enhancement, and multicondition training (MCT) type indicates what kind of multicondition training data was used. All results use per-utterance feature-based maximum likelihood linear regression (fMLLR) for adaptation and minimum Bayes risk (MBR) for decoding. Optional discriminative training is performed using boosted maximum mutual information (bMMI). Optimized decoding refers to optimizing language model weight and beam-width.} \label{table:asr1}
      \begin{tabular}{c | cccc}
        \begin{tabular}{@{}c@{}} Beamforming, \\ TF type, \\ MCT type \end{tabular}   & Clean trained  & MCT  & \begin{tabular}{@{}c@{}} MCT  \\ +bMMI\end{tabular} & \begin{tabular}{@{}c@{}} MCT  \\ +bMMI \\ +optimized decoding\end{tabular}  \\ \hline \\[1.25ex]
        None
                  & 33.21/77.78 & 14.88/34.35 & 11.99/30.50 & 11.31/30.72 \\[2.5ex] \hline \\
        \begin{tabular}{@{}c@{}} 8ch MVDR,  \\ No enh., \\ Noisy MCT \end{tabular}
                  & 16.11/53.64 & 11.01/26.57 & 8.21/24.12 &  7.91/23.91 \\[4ex] \hline \\
        \begin{tabular}{@{}c@{}} 8ch MVDR,  \\ STFT $32$ms, \\ Noisy MCT \end{tabular}
                  & 30.33/63.95 & 14.52/33.63 & 10.10/31.80 &  9.84/32.19 \\[4ex] \hline \\
        \begin{tabular}{@{}c@{}} 8ch MVDR,  \\ STFT $128$ms, \\ Enhanced MCT \end{tabular}
                  & 12.06/40.81 & 9.79/24.91 & 7.63/22.21 &  {\bf 7.31}/{\bf 22.31} \\[4ex] \hline \\
        \begin{tabular}{@{}c@{}} 8ch MVDR,  \\ STFChT $128$ms, \\ Noisy MCT \end{tabular}
                  & 13.95/51.30 & 11.17/30.04 & 10.09/29.94 &  9.74/29.86 \\[4ex] \hline \\
        \begin{tabular}{@{}c@{}} 8ch MVDR,  \\ STFChT $128$ms, \\ Enhanced MCT \end{tabular}
                  & 13.95/51.30 & 10.02/29.34 & 8.34/27.76 &  { 7.96}/{ 27.98} \\[4ex] \hline
      \end{tabular}
\end{table*}

\begin{table*}
\caption{ASR results for REVERB evaluation set using the GMM-HMM Kaldi baseline recognizer by Weninger et al. \cite{weninger_merl/melco/tum_2014}. Same format as table \ref{table:asr1}.} \label{table:asr2}
      \begin{tabular}{c | cccc}
        \begin{tabular}{@{}c@{}} Beamforming, \\ TF type, \\ MCT type \end{tabular}   & Clean trained  & MCT  & \begin{tabular}{@{}c@{}} MCT  \\ +bMMI\end{tabular} & \begin{tabular}{@{}c@{}} MCT  \\ +bMMI \\ +optimized decoding\end{tabular}  \\ \hline \\[1.25ex]
        None
                  & 32.77/77.68 & 15.03/33.96 & 12.45/30.23 & 11.97/30.27 \\[2.5ex] \hline \\
        \begin{tabular}{@{}c@{}} 8ch MVDR,  \\ No enh., \\ Noisy MCT \end{tabular}
                  & 17.50/54.14 & 11.72/25.72 & 8.95/21.96 &  8.82/21.68 \\[4ex] \hline \\      
        \begin{tabular}{@{}c@{}} 8ch MVDR,  \\ STFT $32$ms, \\ Noisy MCT \end{tabular}
                  & 28.49/61.61 & 12.87/29.30 & 10.32/27.13 &  10.14/26.93 \\[4ex] \hline \\
        \begin{tabular}{@{}c@{}} 8ch MVDR,  \\ STFT $32$ms, \\ Enhanced MCT \end{tabular}
                  & 12.86/41.38 & 10.29/22.34 & 7.84/19.71 &  {\bf 7.62}/{\bf 19.84} \\[4ex] \hline \\
        \begin{tabular}{@{}c@{}} 8ch MVDR,  \\ STFChT $128$ms, \\ Noisy MCT \end{tabular}
                  & 14.61/46.70 & 11.54/27.89 & 10.01/24.23 &  9.86/23.99 \\[4ex] \hline \\
        \begin{tabular}{@{}c@{}} 8ch MVDR,  \\ STFChT $128$ms, \\ Enhanced MCT \end{tabular}
                  & 14.61/46.70 & 10.06/25.34 & 8.35/22.77 &  { 8.18}/{ 22.21} \\[4ex] \hline
      \end{tabular}
\end{table*}

\bibliographystyle{ieeetran}
\bibliography{reverb_arxiv}      
\end{document}